\documentclass[a4paper,11pt]{article}
\pdfoutput=1 % if your are submitting a pdflatex (i.e. if you have
\usepackage{jheppub} % for details on the use of the package, please
\usepackage{subfigure}

\usepackage[T1]{fontenc} % if needed

\title{\boldmath Development and Study of a Micromegas Pad-Detector for High Rate Applications}

\author[a]{T.H. Lin, A. D\"udder, M. Schott\footnote{corresponding authors}, C. Valderanis}
\affiliation[a]{Johannes Gutenberg-University, Mainz, Germany}

\emailAdd{mschott@cern.ch}

\abstract{
In this paper, the design and the performance of two prototype detectors based on Micromegas technology with a pad readout geometry is discussed. In addition, two alternative implementations of a spark-resistent protection layer on top of the readout pads have been tested to optimize the charge-up behavior of the detector under high rates. The prototype detectors consist of 500 pads with a size of $5\times 4\,$mm, each connected to one independent readout channel, and cover an active area of $10\times10\,$cm. The design of these prototypes and its associated readout infrastructure was developed in such a way that it can be easily adapted for large-size detector concepts. 
}

\begin{document} 
\maketitle
\flushbottom
\newpage

\section{Introduction}

\begin{figure}[htb]
\begin{minipage}[hbt]{0.49\textwidth}
	\centering
	\includegraphics[width=0.99\textwidth]{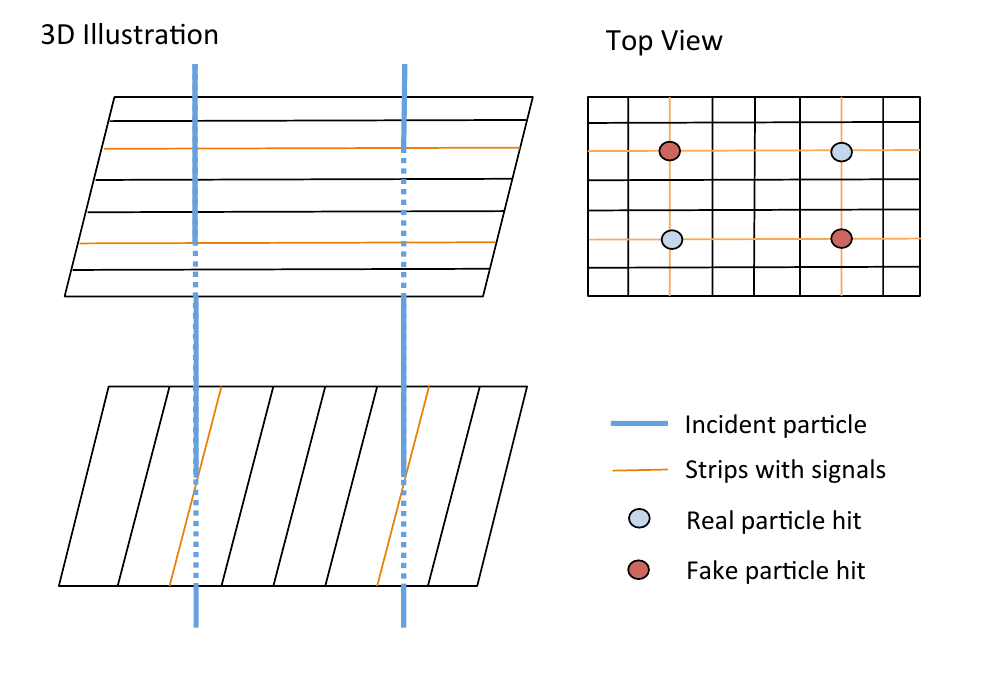}
	\caption{Illustration of combinatorical ambiguities for one-dimensional strip-based detectors.\vspace{0.3cm}}
	\label{Fig:Problem}
\end{minipage}
\hspace{0.2cm}
\hfill
\begin{minipage}[hbt]{0.49\textwidth}
	\centering
	\includegraphics[width=0.99\textwidth]{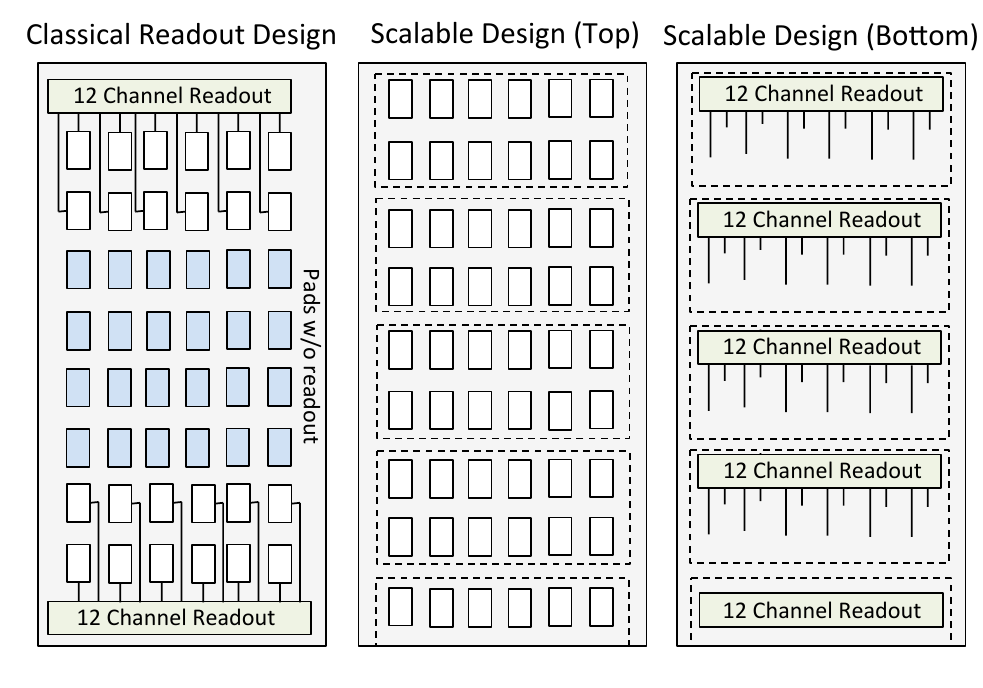}
	\caption{Schematic layout of a classical pad design (left) and scalable pad detector design including readout electronics (right).}
	\label{Fig:DetectorLayout}
\end{minipage}
\end{figure}

Gaseous ionization detectors for particle tracking that aim for the reconstruction of two-dimensional positions are typically based on two separate one-dimensional readout planes that are rotated against each other. The rotation angle between the two layers defines the precision achievable in the reconstruction of the two corresponding coordinates\footnote{e.g. a rotation angle of $90^\circ$ provides an equal precision for the two dimensions}. At high incident-particle rates, combinatorical ambiguities can appear during the reconstruction of the particle position (Figure \ref{Fig:Problem}). In cases where it is not possible to distinguish the recorded signal of two incident particles in two strips in each layer via timing measurements, two so-called 'ghost' solutions appear, as also illustrated in Figure \ref{Fig:Problem}. One solution to this problem is to employ detectors with a pad-readout geometry that allow to solve these combinatorical ambiguities. Other studies of gaseous detectors with pad-geometry, e.g. \cite{Adloff:2013wea}, mainly focussed on the application for calorimeters and hence their readout pad design is based on relatively small pad densities and large distances between individual pads. This has to be different for tracking purposes in high rate environments, which have to accommodate a high pad density including readout electronics on a small surface. We present here for the first time a Micromegas-based pad-detector for application in high-rate environments and large-scale tracking detectors. 

A detailed introduction to the basic operation principle of a Micromegas detector can be found in the literature \cite{Giomataris:1995fq, Alexopoulos:2010zz} and hence we give here only a brief summary. Micromegas are gaseous parallel-plate detectors with two regions that are separated from each other by a thin metallic mesh. The 'drift region' has a typical height of a few mm in which the traversing charged particles ionise gas atoms. The resulting electrons from this ionisation process drift along the electric field lines towards the mesh, which appears transparent to these electrons when the appropriate electric field values are applied. The amplification region has a typical height of 100\,$\mu$m and is next to the separation mesh. The electrical field in the amplification region is higher by two orders of magnitude and is therefore large enough to create an electron avalanche, leading to an amplification of the electron signal by a factor of $\sim10^4$ within less than a nanosecond. These secondary electrons together with the movement of the corresponding positive ions induce a signal on the readout electrodes. 

The large electric fields in the amplification region can lead to sparks, resulting in a dead-time and potential damage to the detector and
the subsequent front-end electronics. A resistive protection layer is therefore typically deposited on top of the readout strips \cite{Alexopoulos:2011zz}. The concept of such insulating protection layers is addressed in more detail in the next sections, as we have developed and studied two alternative implementations in order to optimize the charge-up behavior of the detector under high rates.

The paper is structured as follows. In section \ref{sec:pad}, the scalable design of the micromegas pad detectors and also the two options for the resistive protection layers are discussed. The basic signal shapes and efficiencies for both prototype detectors are discussed in section \ref{sec:perf}, while the performance studies under high rates are summarized in section \ref{sec:rates}. The paper concludes with a summary and an outlook in section \ref{sec:concl}.

\section{\label{sec:pad}Prototype Pad Detectors}

\subsection{\label{sec:design}Scalable Pad Detector Design and Construction}

The basic problem with large-sized gaseous pad detectors is the placement of readout infrastructure, which is typically foreseen at one of the sides. The space available for the readout is limited by the width of the detector and therefore constrains the number of pads since their number, and therefore also the number of readout channels, scales quadratically with the width of the detector (Figure \ref{Fig:DetectorLayout}). In our alternative concept, we place the readout infrastructure on the opposite side of the readout PCB and group the pads such that their covered area corresponds to the required space of the readout-chips, as also illustrated in Figure \ref{Fig:DetectorLayout}. For this, each pad has to be connected through the PCB to the opposite side to be read out. The advantage of this design lies in its scalability, i.e. the possibility to create large active detector volumes, without the need to foresee a readout infrastructure at the sides of the detector. 

\begin{figure}[htb]
\begin{minipage}[hbt]{0.49\textwidth}
	\centering
	\includegraphics[width=0.99\textwidth]{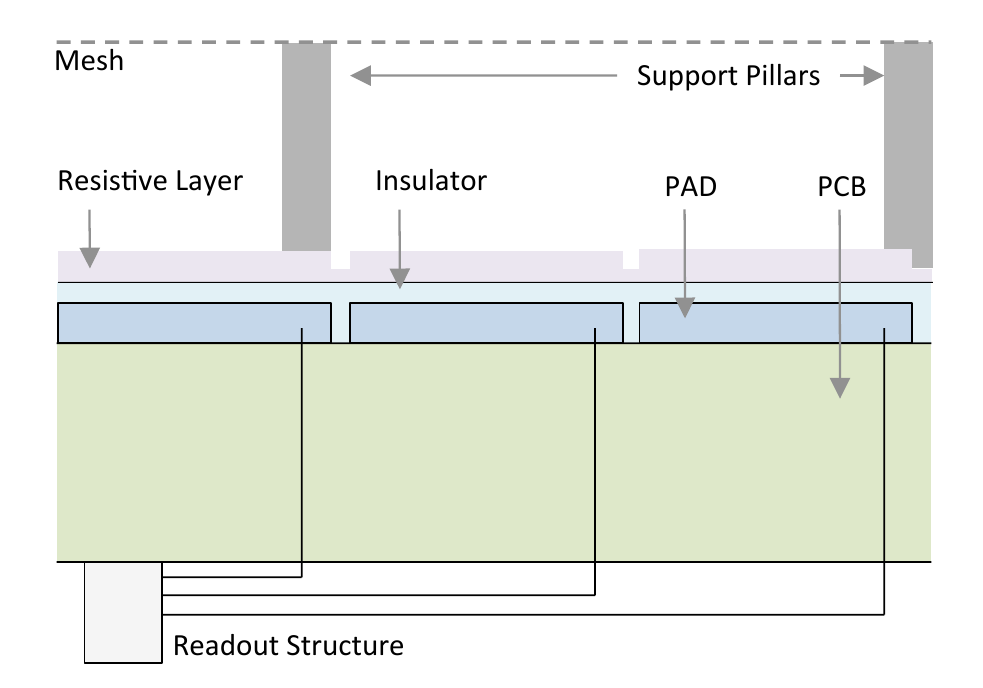}
	\caption{Schematic drawing of the side view on the pad prototype detectors.}
	\label{Fig:ScetchDetectors}
\end{minipage}
\hspace{0.2cm}
\hfill
\begin{minipage}[hbt]{0.49\textwidth}
	\centering
	\includegraphics[width=0.99\textwidth]{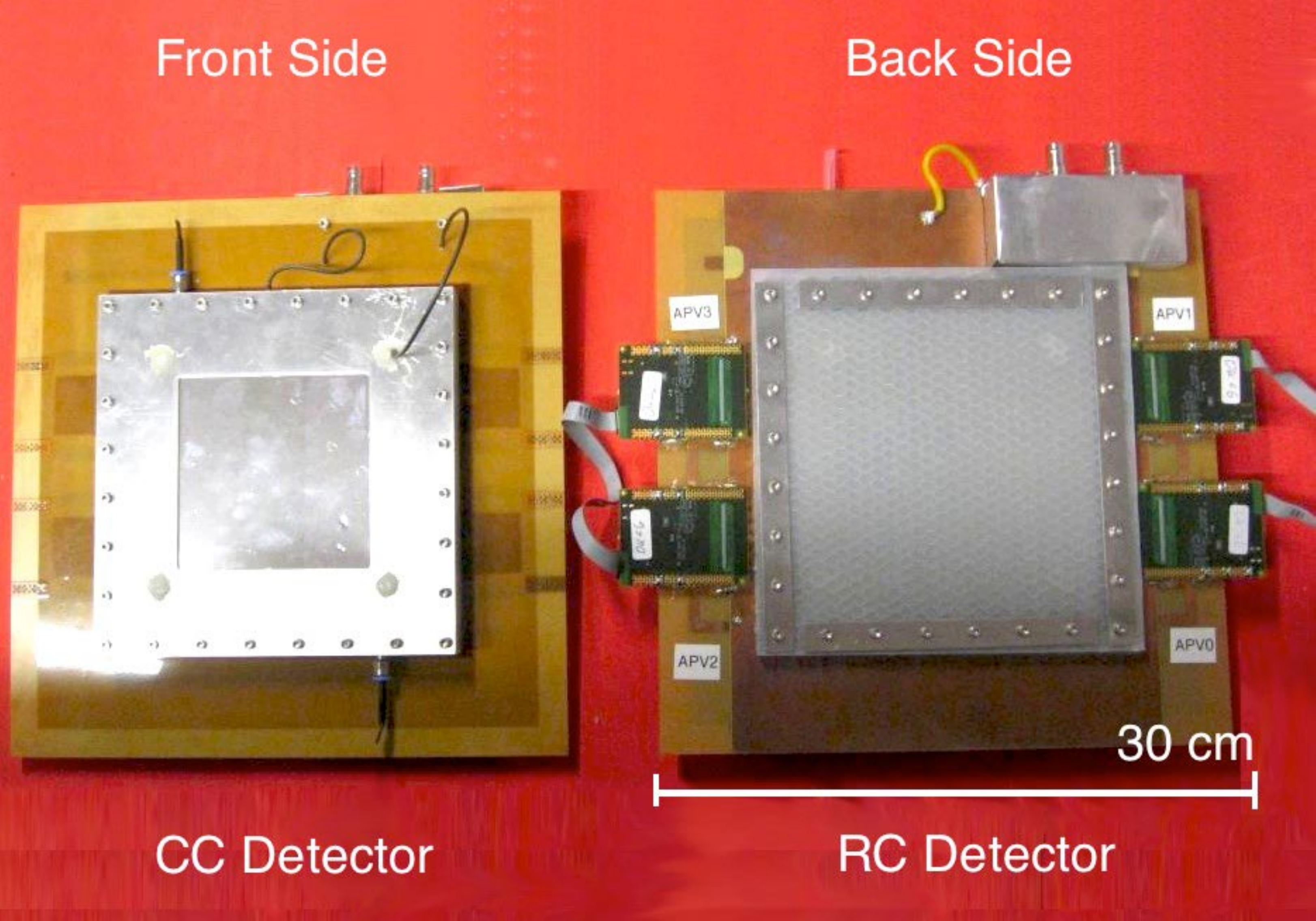}
	\caption{Picture of both prototype detectors from the front and the back side.}
	\label{Fig:Detectors}
\end{minipage}
\end{figure}

Our two prototype detectors follow this design idea and comprise 25 pads in x- and 20 pads in y- direction, leading in total to 500 pads arranged equidistantly in an active area of $10\times10$ cm. The pad size is $5 \times 4\,$mm with a distance of $\approx 50\mu m$ between different pads. A resistive protection layer has been placed on the readout pads, where the implementation for this layer differs for the two prototype detectors (Section \ref{sec:Cap}). The height of the drift and amplification regions is $5\,$mm and $128\,\mu m$ for both prototype detectors, respectively. The amplification mesh is made of stainless steel with a density of 157 lines/cm and a line diameter of $25\,\mathrm{\mu m}$ and covers the full active area. Its distance to the readout pads is guaranteed by support pillars that have a $0.4\,\mathrm{mm}$ diameter. The support pillars are placed along a regular matrix with $2.5\,$mm spacing in both directions. We have chosen a floating mesh assembly approach for mounting the amplification mesh, i.e. we first glued the amplification mesh onto a support frame and then attached the support frame face-down to the readout board. The floating mesh assembly strategy often leads to a bending of the small readout PCBs and therefore implies a non-uniform height between amplification mesh and readout plane. To avoid these non-uniformities, an additional support structure was mounted onto the bottom side of the readout-board. A scetch of the prototype detector design is illustrated in Figure \ref{Fig:ScetchDetectors}, while a picture of the final constructed prototypes is shown in Figure \ref{Fig:Detectors}.

\subsection{\label{sec:Cap}Capacitively and Resistively Coupled Spark Resistant Protection Layers}

\begin{figure}[htb]
\begin{minipage}[hbt]{0.49\textwidth}
	\centering
	\includegraphics[width=0.99\textwidth]{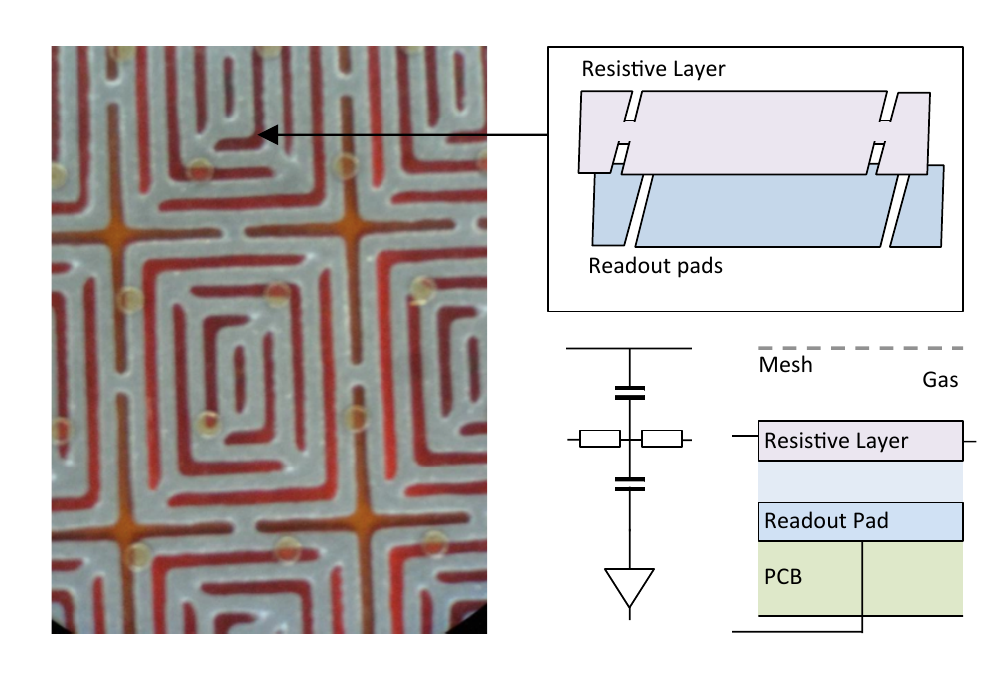}
	\caption{Picture of the resistively pads for the capacitive-coupled detector layer including a readout circuit.}
	\label{Fig:CapacitiveCoupled}
\end{minipage}
\hspace{0.2cm}
\hfill
\begin{minipage}[hbt]{0.49\textwidth}
	\centering
	\includegraphics[width=0.99\textwidth]{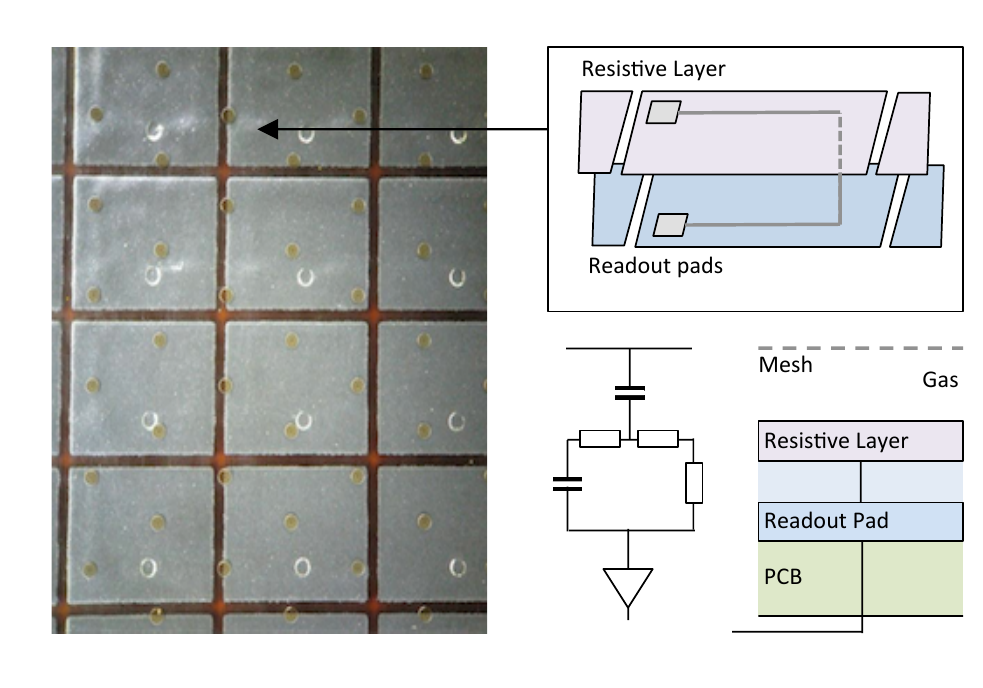}
	\caption{Picture of the resistively pads for the resistive-coupled detector layer including a readout circuit.}
	\label{Fig:ResistiveCoupled}
\end{minipage}
\end{figure}

In spark-resistant micromegas detectors with a strip-readout design, the resistive strips are directly covered by other resistive strips with a resistivity of $\sim 20\,\mathrm{M\Omega/cm}$ \cite{Alexopoulos:2011zz, Lin:2014jxa}. In this case, the resistive strips are grounded at the sides of the detector and the signal is induced via capacitive coupling onto the readout strips. We have chosen a similar approach for the first prototype detector. The resistive layer has also a pad structure, i.e. each readout pad is covered individually by one corresponding resistive pad. In addition, all resistive pads are interconnected as illustrated in Figure \ref{Fig:CapacitiveCoupled} so that they can be commonly grounded on one side of the detector. Each resistive pad is of a labyrinth-like shape, which was chosen to reduce the charge spread to the neighboring pads. The resistance between two pads is $\approx 5-10\,\mathrm{M\Omega}$. It should be noted that the principle of this design is similar to typical resistive strip designs: the resistive readout layer can be grounded and the electron signal propagates to the sides of the detector where the grounding is applied. The actual signal in the readout pads is therefore induced via capacitive coupling (CC) mainly by the charges that are induced by the ion-drift. We label this prototype detector therefore $CC$-prototype detector in the following.

A potential disadvantage of the CC design becomes apparent at high incident-particle rates and large detectors. First, we expect a significant induced charge also in neighboring pads due to the charge spread via the interconnecting resistive pads. Secondly, we expect a current from the electron avalanche to the ground of the detector at its sides, leading to potential interferences at high rates. Therefore we have chosen to implement an alternative spark protection design as shown in Figure \ref{Fig:ResistiveCoupled}. Here, we connect each resistive pad with the readout pad on the PCB via a high-resistive connection. The resistance between the protection layer and the readout pad is $\approx 5\,\mathrm{M\Omega}$. Hence the individual pads are not connected with each other any more. The signal on the readout strip is induced therefore not only via capacitive coupling, but also via a direct, high-resistive connection between the protection and the readout pad. A schematic circuit of this 'resistively-coupled' (RC) approach is also illustrated in \ref{Fig:ResistiveCoupled}. With this design, we expect only a limited effect between neighboring pads and we also do not expect any currents over the full resistive protection layer. It should be noted, that the readout layer is grounded in the RC-approach while typically the amplification mesh is set to ground for CC-based micromegas detectors. 

It turned out during detector assembly that the edges of the resistive pads are not totally flat but stick out by several $\mu$m. Therefore, we expect a different field-line configuration between the RC and CC detectors, which implies a necessity of a careful choice of the applied amplification voltages to allow a direct performance comparison of both detectors.

\subsection{Experimental Setup and Readout Infrastructure}

The performance of the two prototype detectors was studied for different amplification ($V_A$) and drift voltages ($V_D$) using a 93:7 Ar:$\mathrm{CO_2}$ gas mixture. The amplification voltage is defined as the voltage difference between the readout layer and the amplification mesh, while the drift voltage is defined as the difference between the amplification mesh and the drift electrode. We used a cosmic ray measurement facility to test the detector performance for incident muons and a tunable X-Ray tube (Amptek Mini-X-Ray) to study the detector behavior under high rates.

The initial performance studies are based on a Multi-Channel Analyzer (MCA8000D, Amptek) and a dedicated preamplifier. The following detailed studies of the detector performance are based on the RD51 Scalable Readout System (SRS) \cite{Martoiu:2013aca, Martoiu:2011zja} for the data acquisition. The signal processing of the detector is based on the Analog Pipeline Voltage chip with $0.25\,\mathrm{\mu}$m CMOS technology (APV25) \cite{APVJones}, where the analog signal data is transmitted via HDMI cables to SRS electronics. The SRS electronics processes the analog signal which is then further analyzed. The height of the readout signal recorded (in units of ADC counts) is related to the charge induced on the readout strip. The APV25 integrates the current induced for $75\,$ns, so the resulting charge is only part of the total induced charge. In order to study the actual signal shape before the processing of the APV25 chip-set, we also performed several tests based on a multi-channel analyzer. The corresponding results are discussed in the following sections.

\section{\label{sec:perf}Results of Basic Performance Studies}

\subsection{Basic Signal Shapes}

In a first step, we tested both detectors with the X-ray tube and an $Fe^{55}$ source and recorded the corresponding pulse-height spectra via an MCA-unit applying $V_D=300$ and $V_A=465\,$V on the CC detector and $V_D=300$ and $V_A=450$\,V on the RC detectors, respectively. As already mentioned in Section \ref{sec:Cap}, a difference in the electric field in the amplification region is expected and hence the voltages have to be tuned to reach comparable operation conditions. The voltages have been therefore chosen in such a way that the average observed maximal charge is comparable for both detector types.

\begin{figure}[thb]
    \begin{center}
        \includegraphics[width=0.49\textwidth]{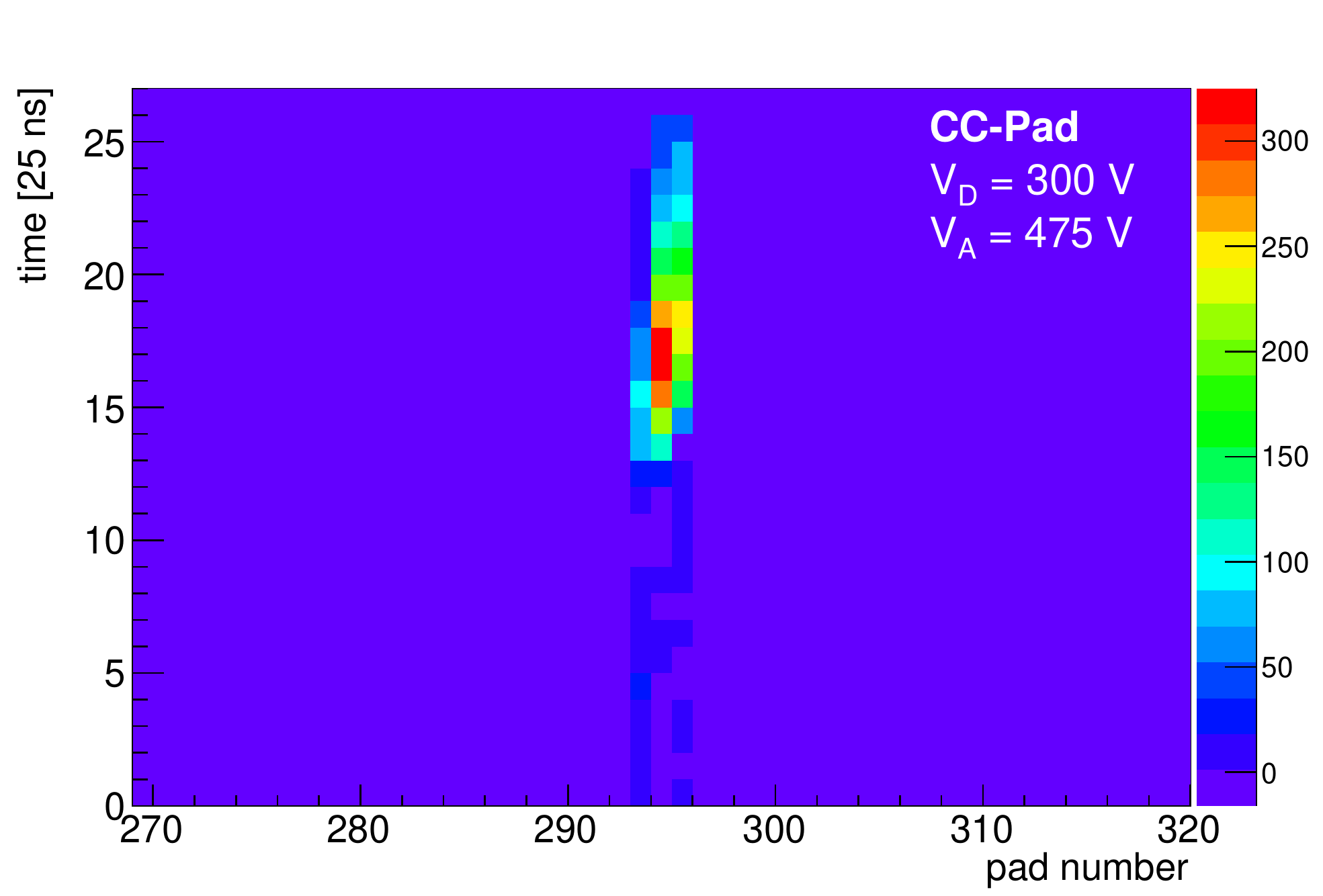}
        \includegraphics[width=0.49\textwidth]{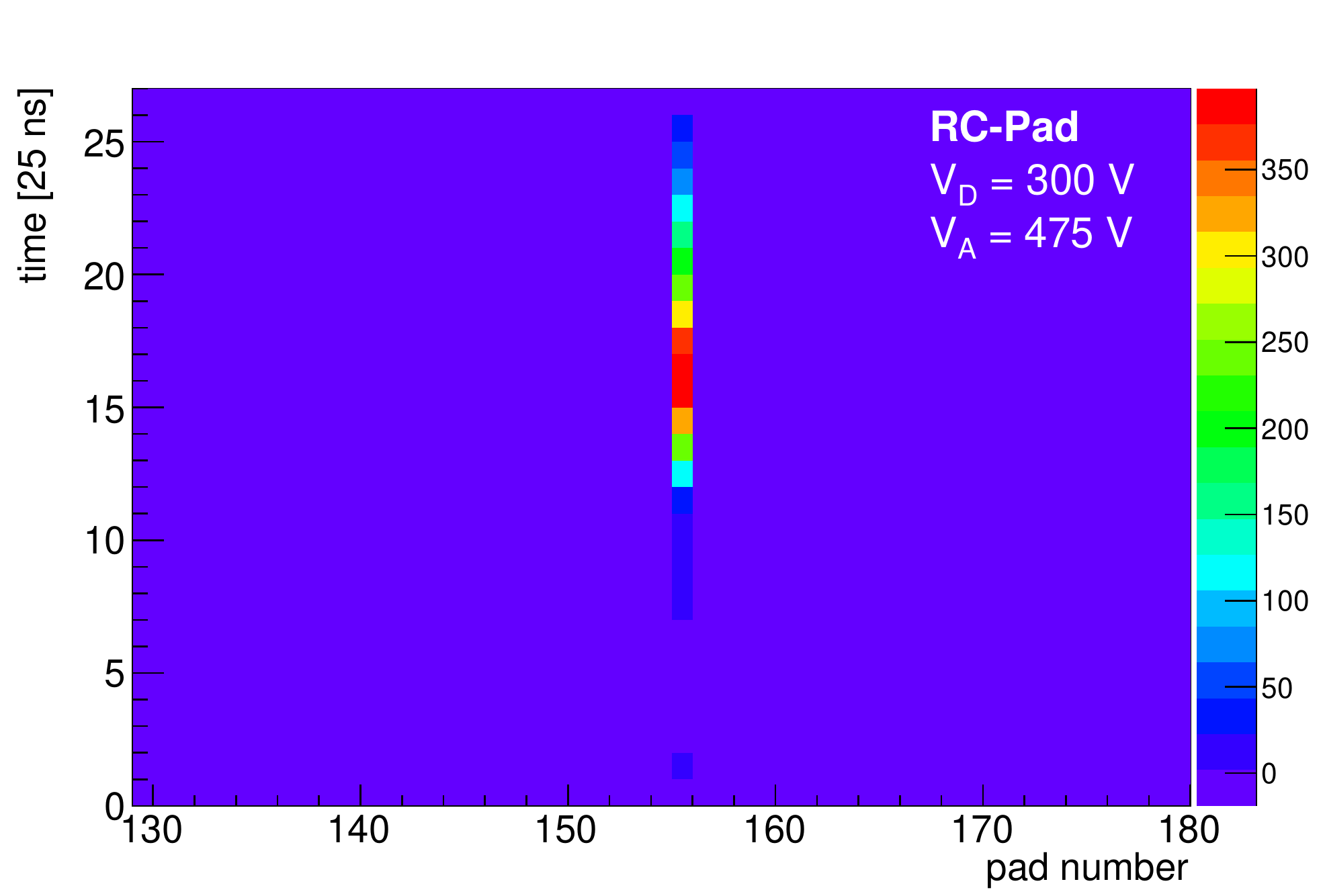}
        \caption{Recorded charge in all pads evolution vs. time for an X-ray event, recorded by the APV-Readout System for the CC (left) and the RC (right) detector.}
        \label{Fig:SignalEvolution}
    \end{center}
\end{figure}

\begin{figure}[thb]
    \begin{center}
        \includegraphics[width=0.49\textwidth]{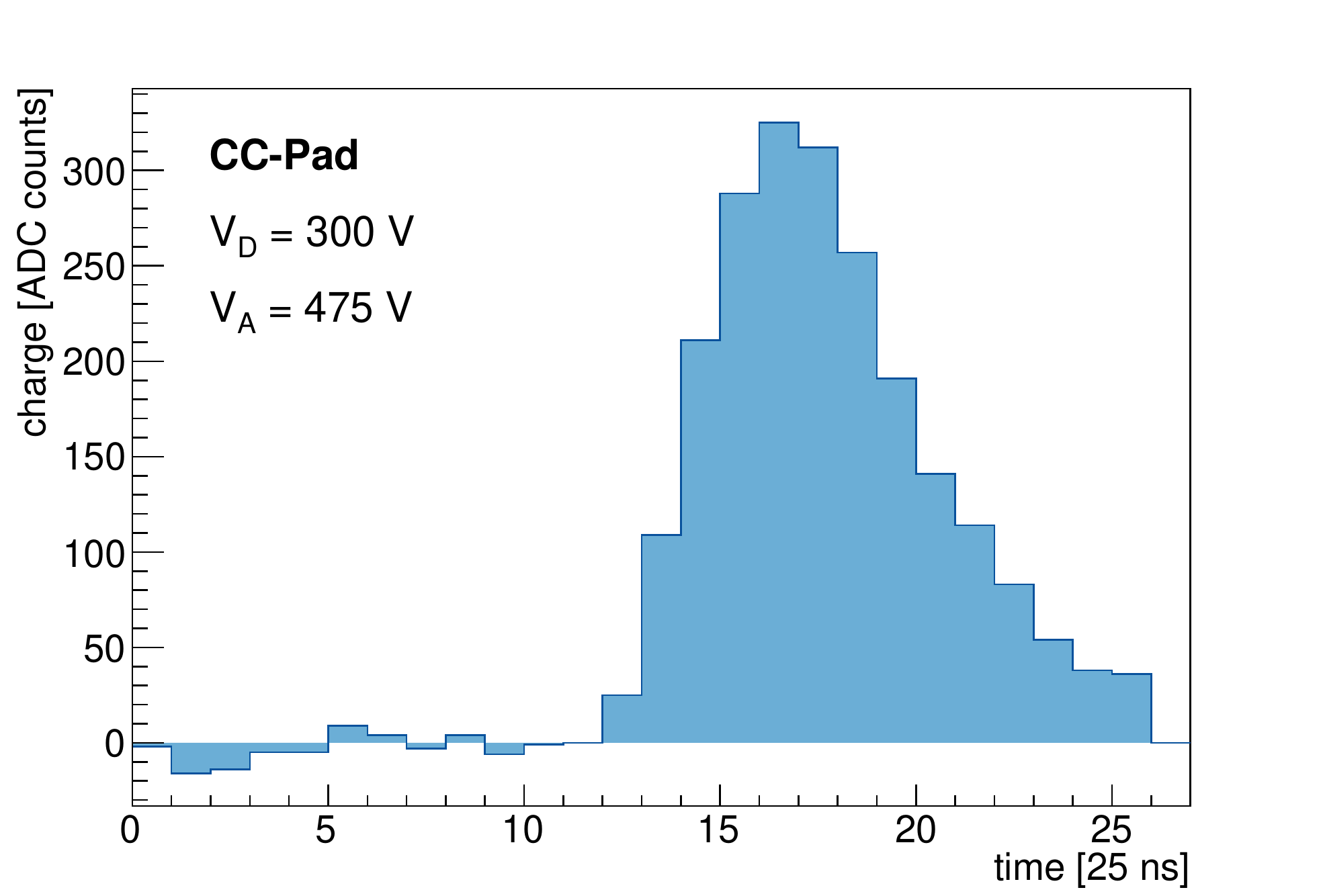}
        \includegraphics[width=0.49\textwidth]{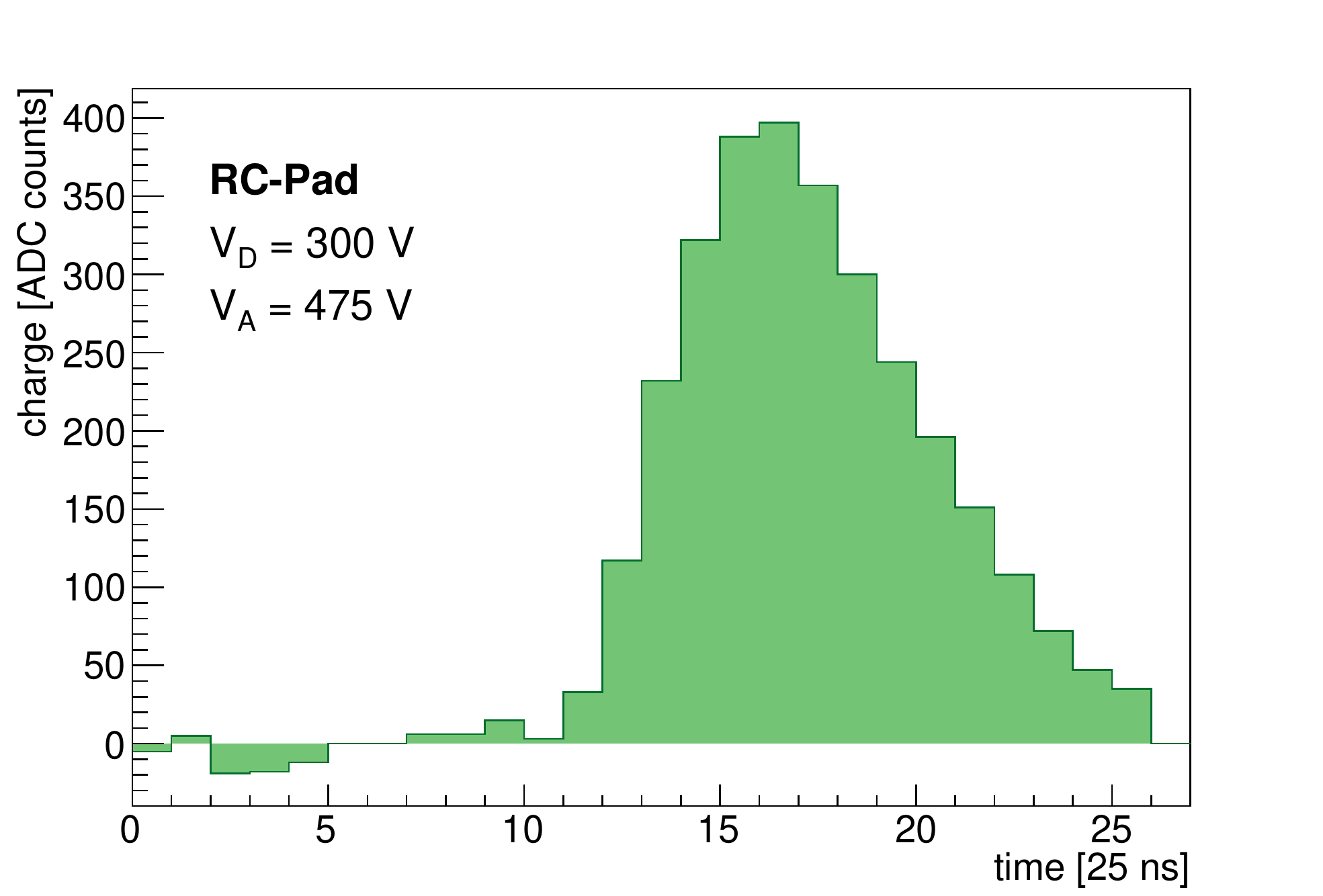}
        \caption{Recorded charge vs. time in the pad containing the maximal recorded charge for an X-ray event, recorded by the APV-Readout System for the CC (left) and the RC (right) detector.}
        \label{Fig:SignalEvolutionInOnePad}
    \end{center}
\end{figure}

A typical recorded signal stemming from X-ray tube is shown for both detectors in Figure \ref{Fig:SignalEvolution}, where the consecutive pad number\footnote{A unique number is assigned to each pad in the two-dimensional pad layout to allow a one-dimensional representation} is shown on the x-axis for 27 time-steps of $25\,$ns length on the y-axis. The signal height recorded on each pad and for each time step is also graphically indicated. A typical event is very localized on one pad and the effect of neighboring pads is typically small. We therefore define a cluster as a set of neighboring pads which all have a recorded signal above a certain threshold (Figure \ref{Fig:SignalEvolution}). The recorded signal height in ADC counts for the pad with the maximal recorded charge $q_{max}$ is shown in Figure \ref{Fig:SignalEvolutionInOnePad} for both detectors. The maximal recorded charge is defined as the time-step which contains the highest ADC count rate. We require a lower threshold on $q_{max}$ of $10$ ADC counts to define a hit in one pad. Furthermore we define the \textit{signal decay time} $\Delta t$ as the time-difference between the maximum of recorded charge distribution to the time-step where the signal height has decreased to 10\% of the maximal recorded charge (Figure \ref{Fig:SignalEvolutionInOnePad}). The signal height, as well as the signal decay time $\Delta t$ are comparable for both detectors for the operation conditions chosen. A more detailed study is performed in the following. 

\subsection{Detector Performance Dependence on Drift- and Amplification-Voltage}

In order to study the performance of both prototype detectors under different operating conditions, an X-ray generator (Amptek Mini-X) was used to generate highly energetic electrons within the drift region via the photoelectric effect. A random trigger was used during the data-taking.

\begin{figure}[thb]
    \begin{center}
        \includegraphics[width=0.49\textwidth]{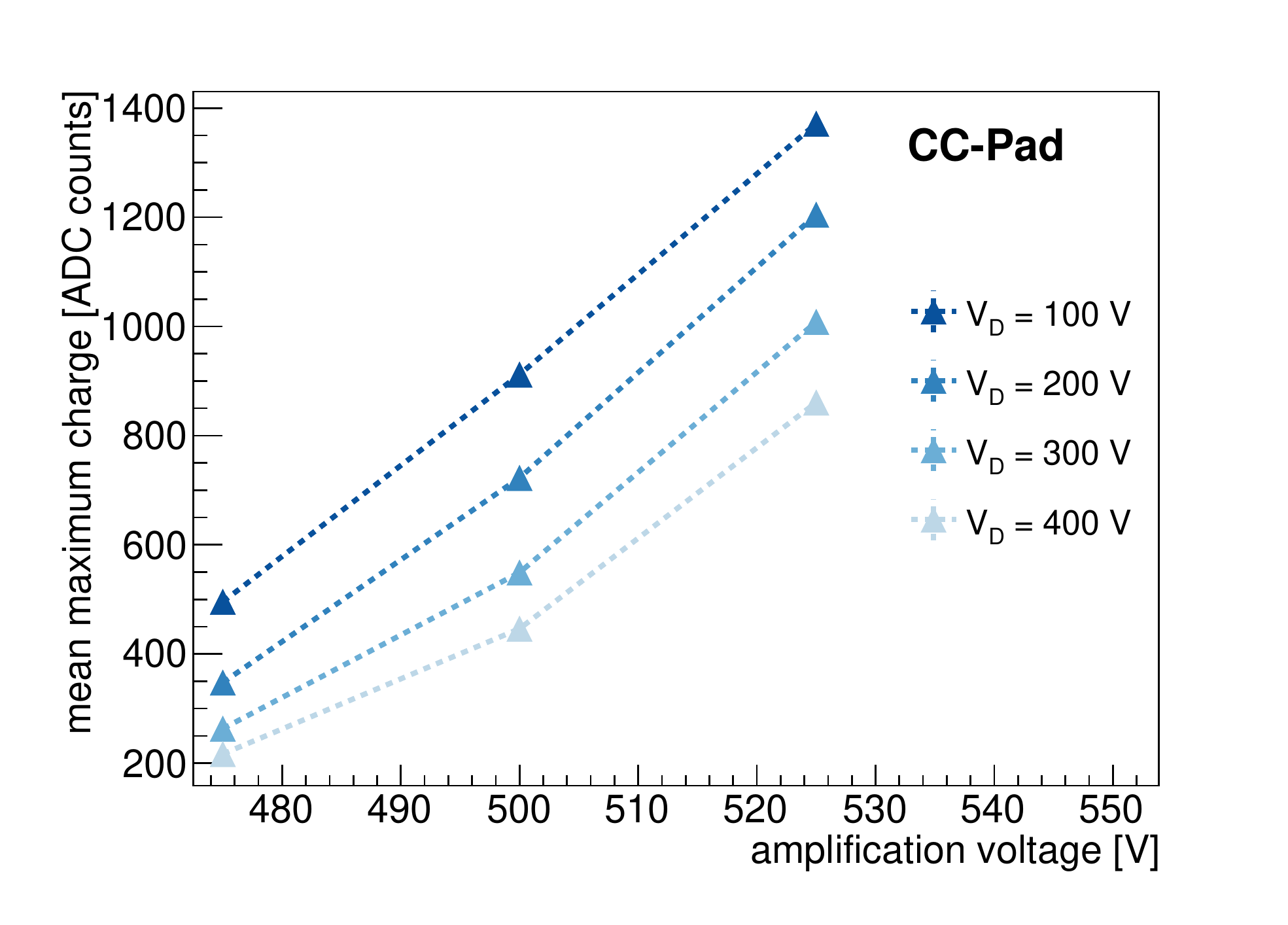}
        \includegraphics[width=0.49\textwidth]{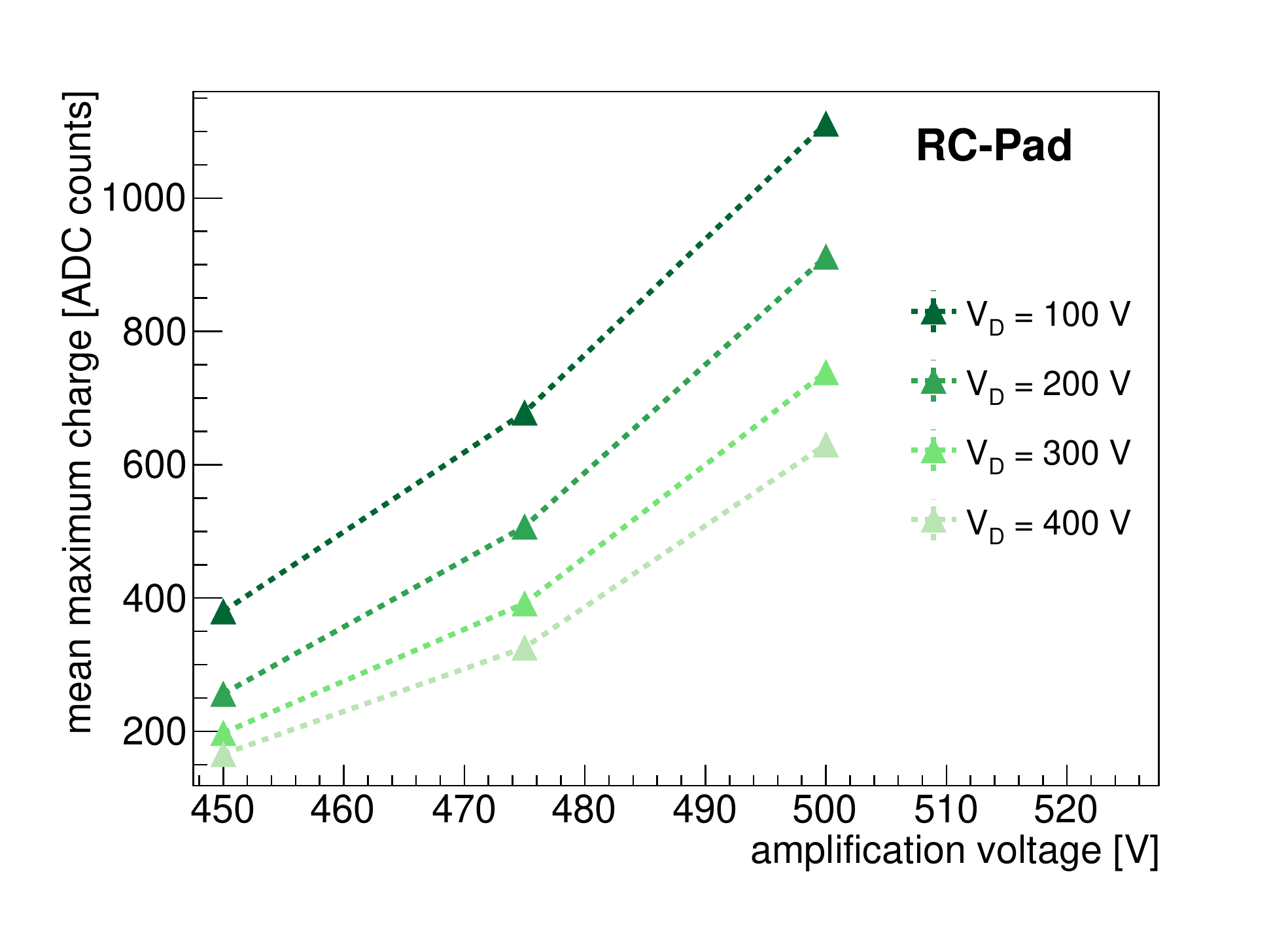}
        \caption{Dependence of the mean maximal recorded charge on the amplification voltage ($V_A$) for different drift voltages ($V_D$) for the CC (left) and the RC (right) detector.}
        \label{Fig:DepAD}
    \end{center}
\end{figure}

\begin{figure}[thb]
    \begin{center}
        \includegraphics[width=0.49\textwidth]{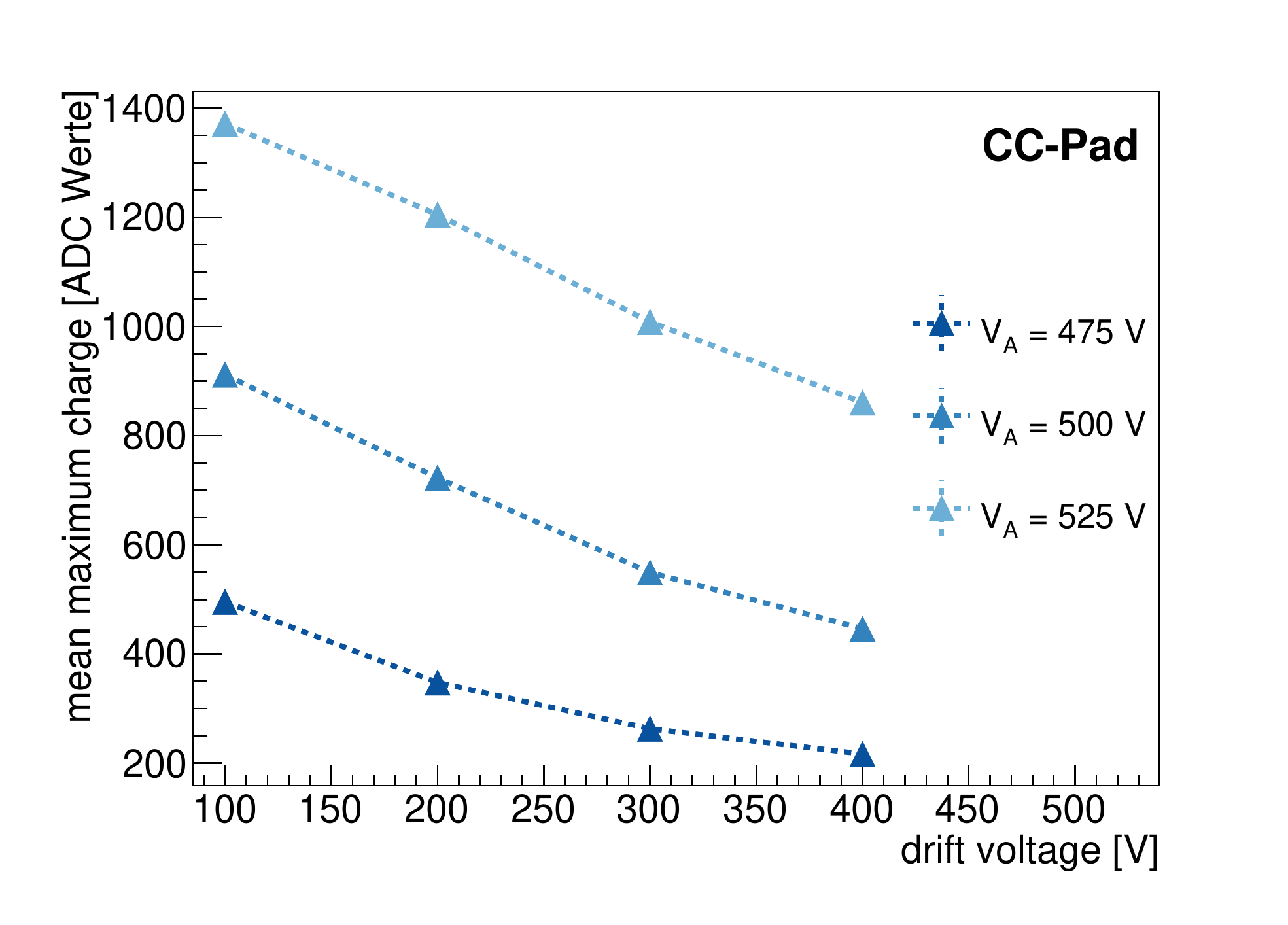}
        \includegraphics[width=0.49\textwidth]{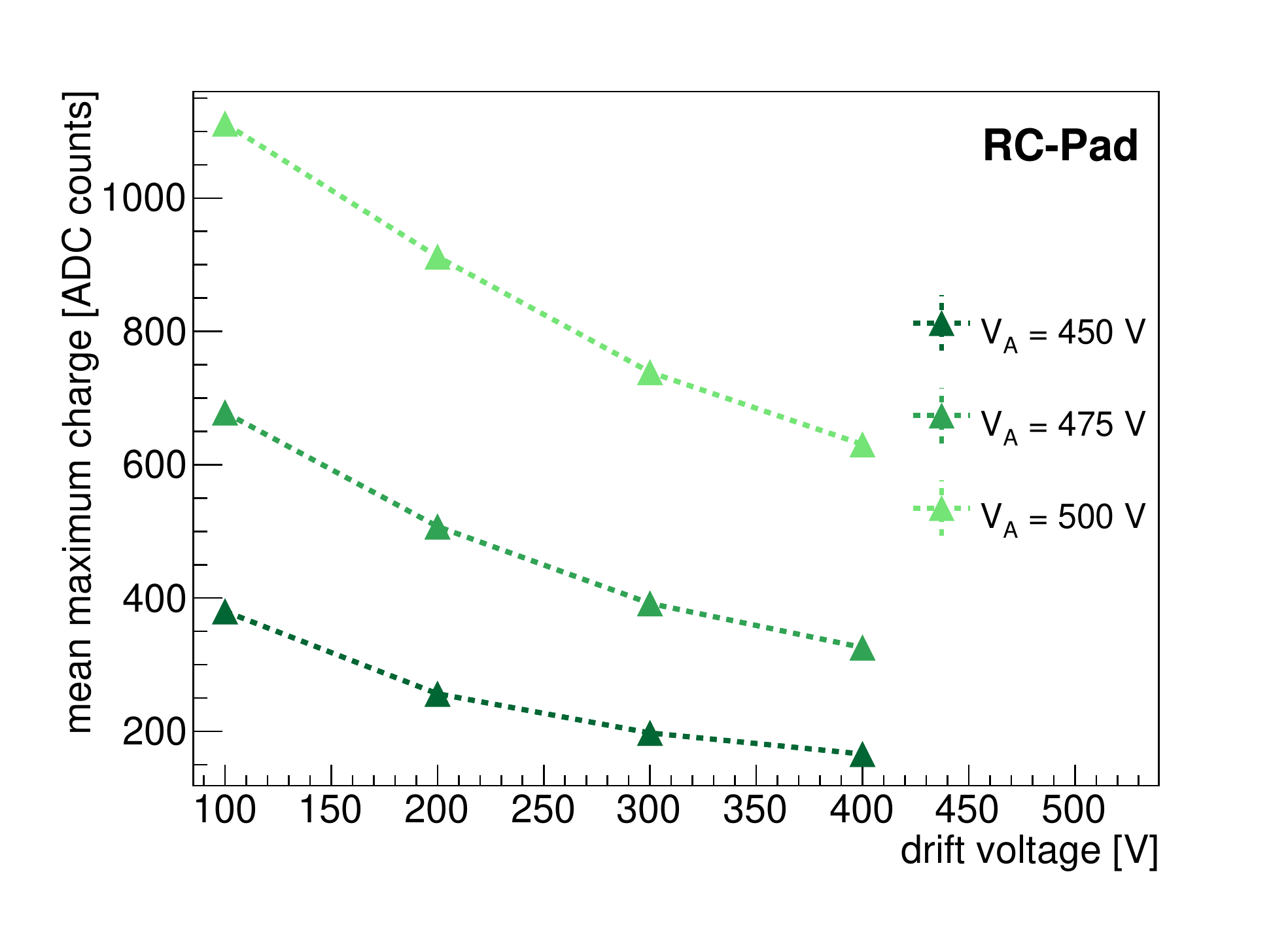}
        \caption{Dependence of the mean maximal recorded charge on the drift voltage ($V_D$) for different amplification voltages ($V_A$) for the CC (left) and the RC (right) detector.}
        \label{Fig:DepVD}
    \end{center}
\end{figure}

The study of the dependence of the maximum recorded charge on the drift- and amplification voltages is performed with such X-ray measurements. The results are shown in Figures \ref{Fig:DepAD} and \ref{Fig:DepVD}, respectively. Even though the nominal values of the maximal recorded charge are different for both detectors, we observe a similar dependence on $V_D$ and $V_A$. The rise of $q_{max}$ vs. $V_A$ is expected, as the electric field in the amplification region rises and leads to an increase of the amplification. A decrease of $q_{max}$ vs. $V_D$ can be observed for both detectors. This is due to a decrease in mesh transparency with higher drift-voltages, i.e. fewer electrons from the drift region reach the amplification region, leading to an overall decrease of the average maximum charge. 

\subsection{Reconstruction Efficiency in Cosmic-Ray Tests}

The reconstruction efficiency of both prototype detectors was tested in a cosmic ray measurement facility. Two micromegas detectors with an active area of $10\times10$cm and a two-dimensional readout \cite{Lin:2014jxa} have been placed above and below the pad prototype detectors, allowing for a tracking of incident cosmic rays. A coincident signal of two photomultipliers has been used as trigger. When signals in both reference chambers could be reconstructed and the corresponding reconstructed track was expected within the active area of the pad detectors, we tested wether a maximal recorded charge $q_{max}$ above the threshold close to the predicted particle impact region has been registered. The resulting efficiencies has been tested in 16 different detector regions and turn out the homogenous for both detectors. An average hit-reconstruction efficiency of $\approx 95 \%$ has been estimated. 

%The resulting efficiencies are shown in Figure \ref{Fig:Efficiency} for 16 detector regions. Again, we observe a very similar behavior for both prototypes with an average efficiency of $\approx 80-85 \%$\footnote{Two neighboring regions in the RC detector have been damaged during the test setup and have been excluded form this study}. 

%\begin{figure}[htb]
%    \begin{center}
%        \includegraphics[width=0.49\textwidth]{Figures/RecoEfficiency_CC.pdf}
%        \includegraphics[width=0.49\textwidth]{Figures/RecoEfficiency_RC.pdf}
%        \caption{Signal reconstruction efficiency for cosmic-ray muons for the CC (left) and the RC (right) detector.}
%        \label{Fig:Efficiency}
%    \end{center}
%\end{figure}
%
%

\section{\label{sec:rates}Detector Performance under High Rates}

As discussed in the previous section, the basic performance observables showed no significant difference between the two different pad detector prototypes. In this chapter, we investigate the performance behavior of both detectors in dependence of the incident-particle rates with the Amptek X-ray source with which the current of the X-ray tube, and therefore the photon rate, can be tuned. 

\begin{figure}[tb]
\begin{minipage}[hbt]{0.49\textwidth}
	\centering
	\includegraphics[width=0.99\textwidth]{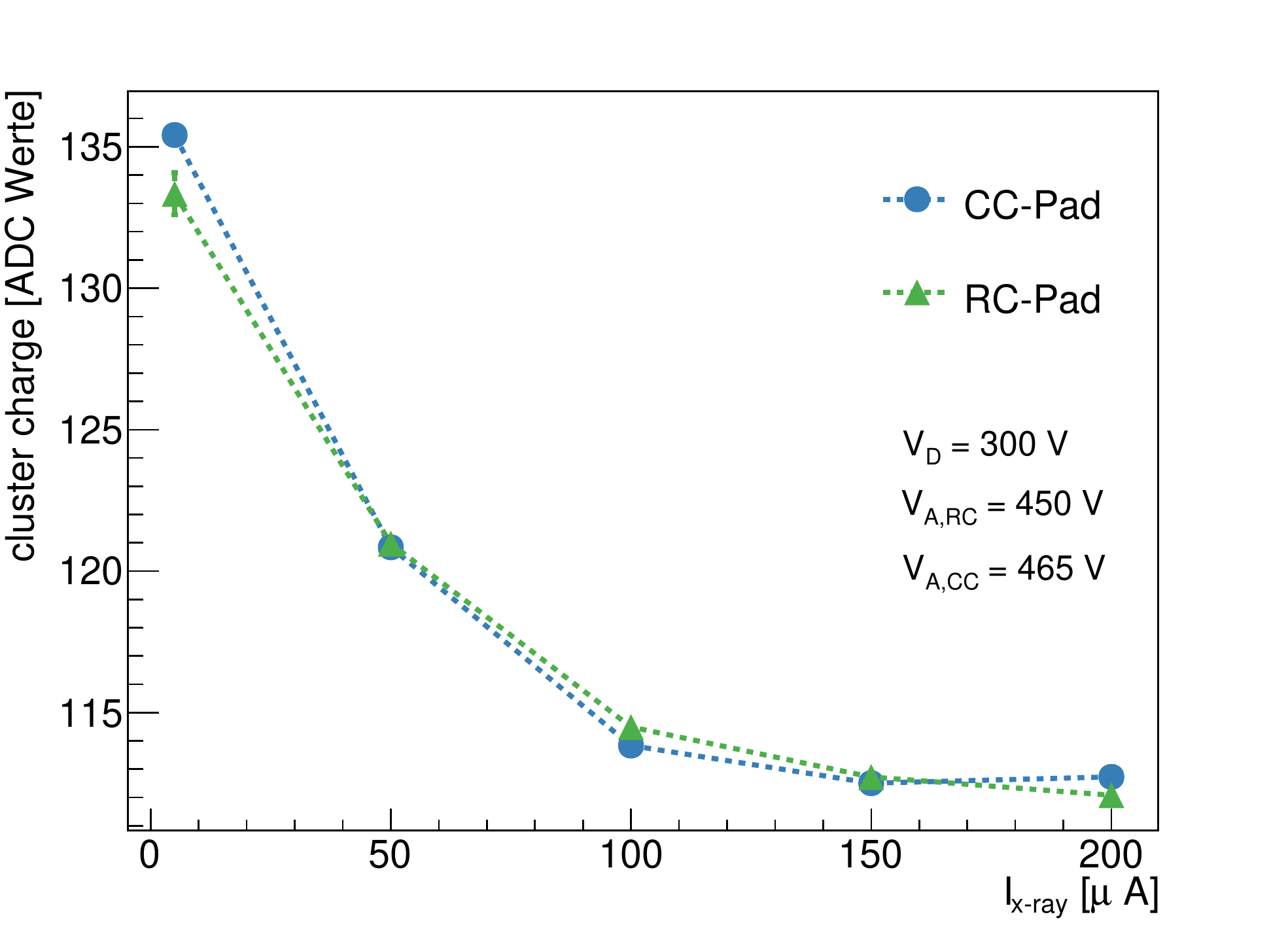}
	\caption{Average charge per cluster in dependence of the current at the X-Ray source for the CC and the RC detector.}
	\label{Fig:AvChargeCurrent}
\end{minipage}
\hspace{0.2cm}
\hfill
\begin{minipage}[hbt]{0.49\textwidth}
	\centering
	\includegraphics[width=0.99\textwidth]{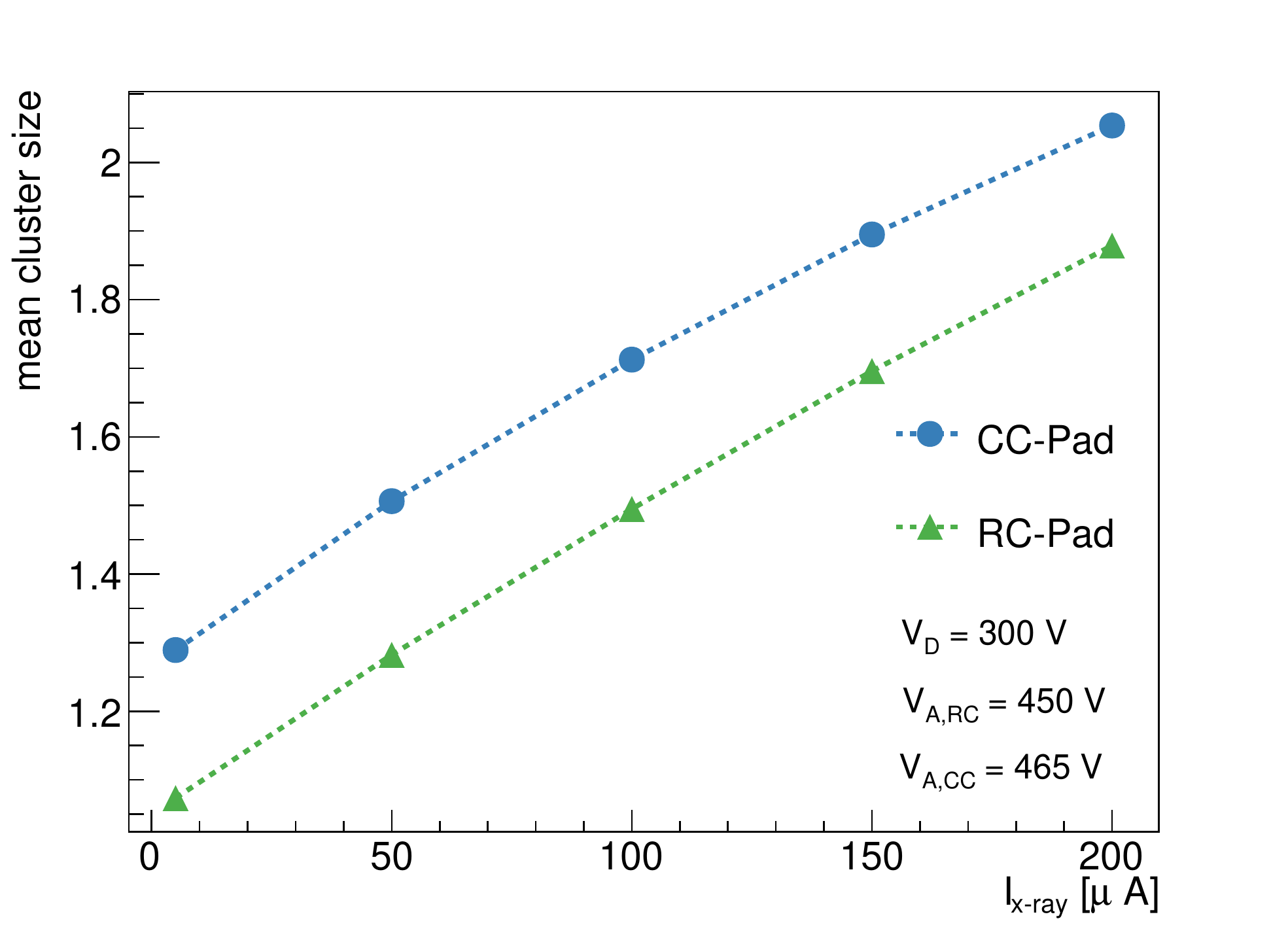}
	\caption{Average number of pads per cluster in dependence of the current at the X-Ray source for the CC and the RC detector.}
	\label{Fig:AvPadCurrent}
\end{minipage}
\end{figure}

Figure \ref{Fig:AvChargeCurrent} shows the dependence of the average cluster charge, i.e. the overall recorded charge in one cluster, on the current at the X-Ray source for the CC and the RC detectors. Both detectors show a very similar behavior, i.e. a decreasing cluster charge with higher rates. This is a result of the screening effect of the  charges produced in the amplification region of the electric field. In higher-rate environments, we expect more charges in the amplification volume, which effectively lower the effective field strength and hence lead to smaller amplifications.

Figure \ref{Fig:AvPadCurrent} shows the average number of pads that are associated to one reconstructed cluster in dependence on the current at the X-Ray source for both detectors. We observe a similar dependence on the current, i.e. on the incident flux rate, but a significantly smaller average number of pads per cluster for the RC detector. This confirms the expected behavior of the initial detector design, as the signal does not spread to neighboring pads and therefore allows for a significantly enhanced signal reconstruction in high-rate environments. 

\begin{figure}[tb]
    \begin{center}
        \includegraphics[width=0.49\textwidth]{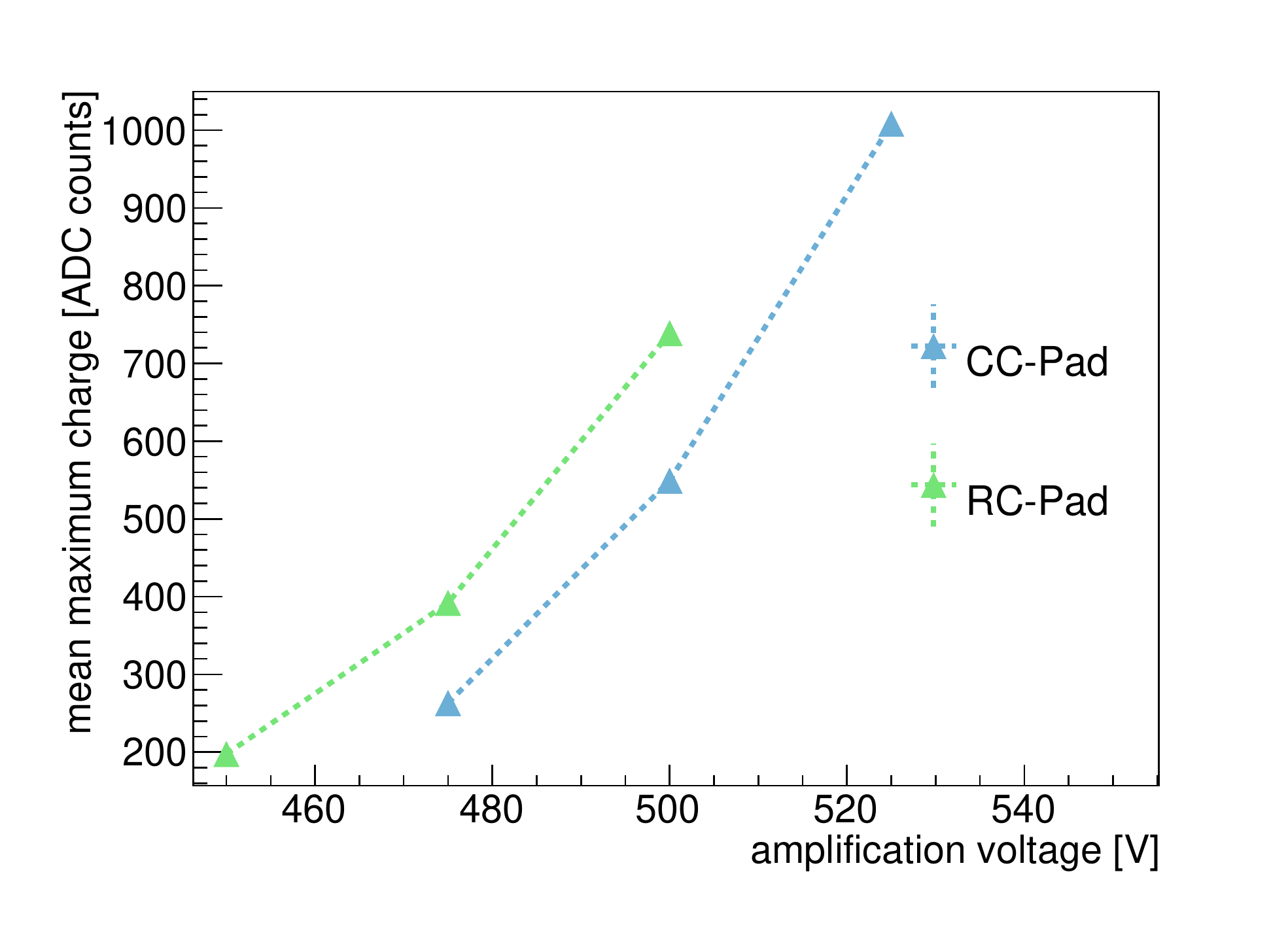}
        \includegraphics[width=0.49\textwidth]{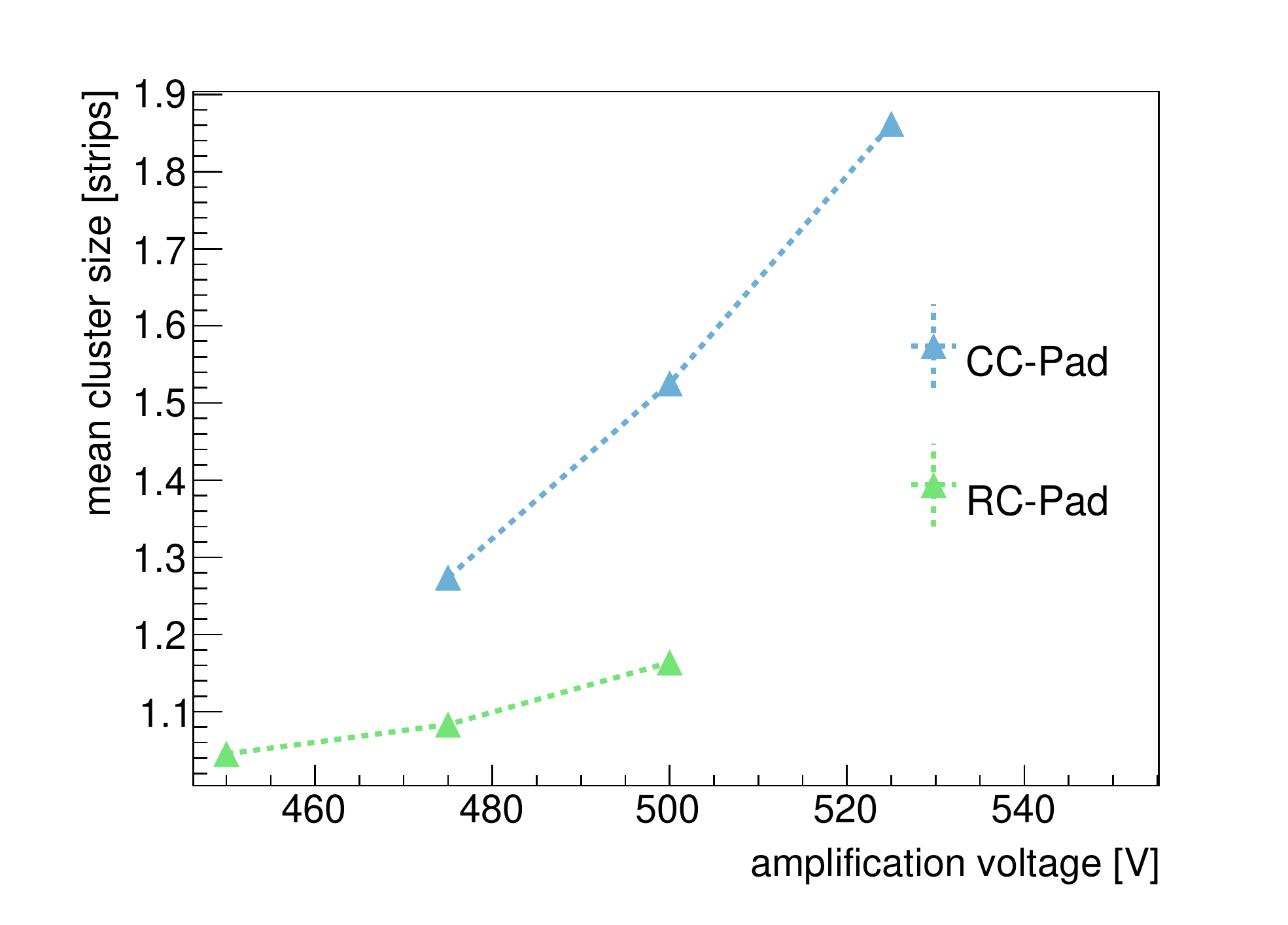}
        \caption{Average maximum charge per cluster (left) and average number of pads per cluster (right) in dependence of $V_A$ for the CC and the RC detector.}
        \label{Fig:SigPadVA}
    \end{center}
\end{figure}

The dependence of the average cluster charge and the cluster-size on the applied amplification voltage $V_A$ is shown in Figure \ref{Fig:SigPadVA}. In particular, the RC detector shows nearly no dependence of the average cluster size on $V_A$, while the cluster size increases by more than $70\%$ for the CC detector. This again highlights the advantages of the direct coupling to the readout pads for the RC detector and the related reduced spread in charges.

\begin{figure}[htb]
\begin{minipage}[hbt]{0.49\textwidth}
	\centering
	\includegraphics[width=0.99\textwidth]{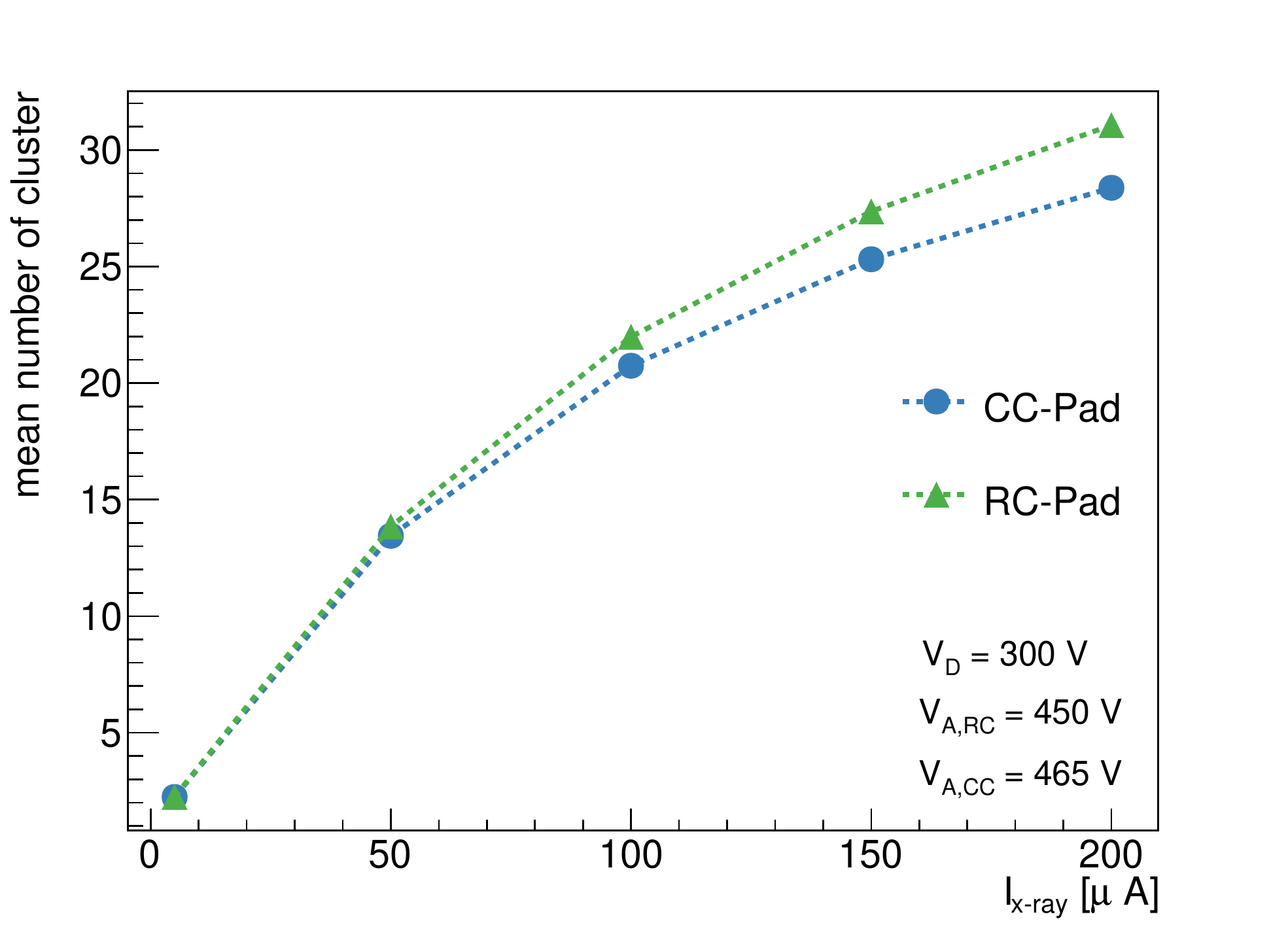}
	\caption{Average number of clusters per one recorded event in dependence on the current at the X-Ray source.}
	\label{Fig:AvClusCurrent}
\end{minipage}
\hspace{0.2cm}
\hfill
\begin{minipage}[hbt]{0.49\textwidth}
	\centering
	\includegraphics[width=0.99\textwidth]{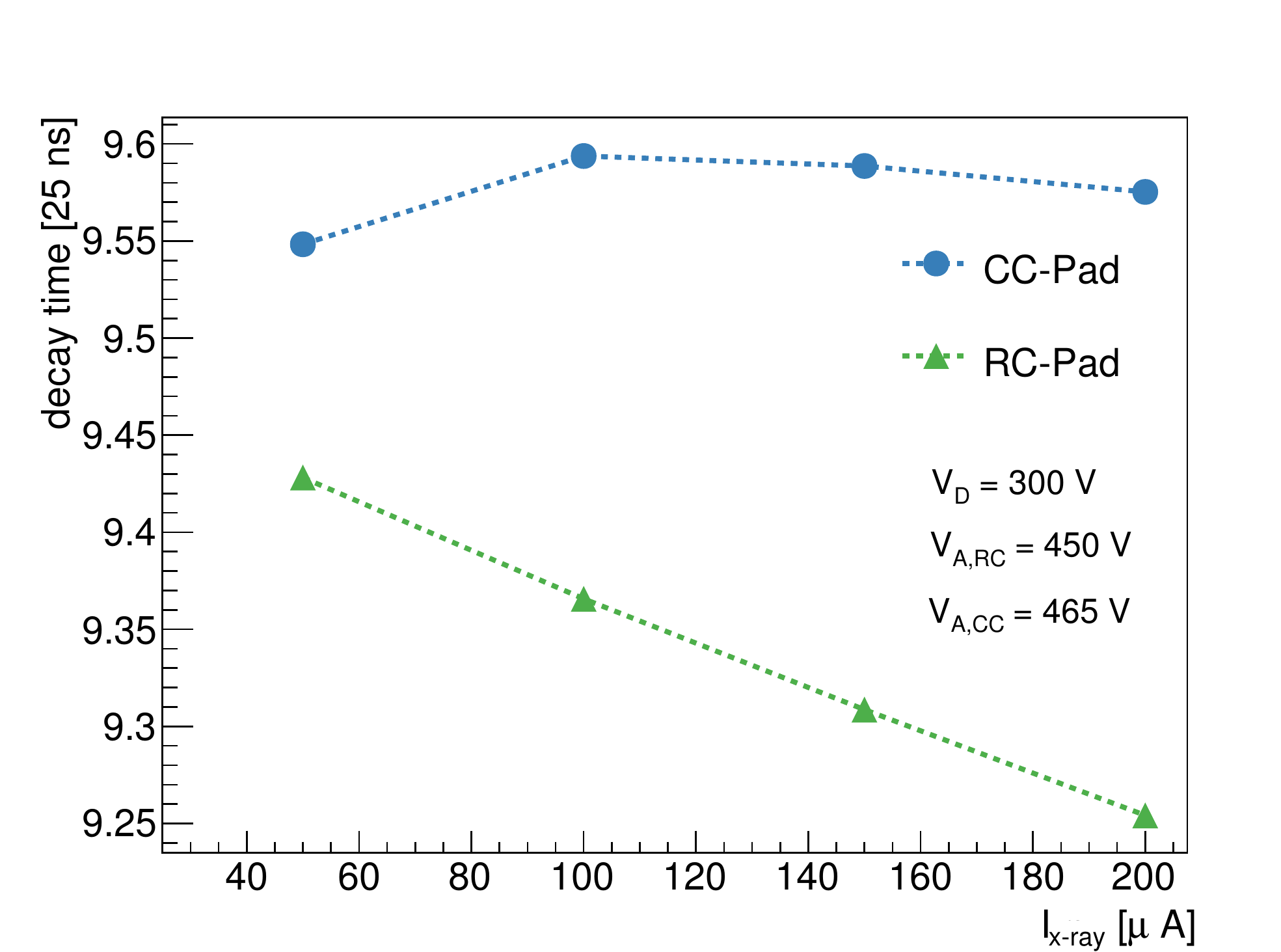}
	\caption{Average signal duration from signal peak to $10\%$ signal height in dependence on the current at the X-Ray source.}
	\label{Fig:SigDurCurrent}
\end{minipage}
\end{figure}

A higher incident flux implies also a higher number of reconstructed clusters per event. The dependence of the average number of clusters on the current at the X-Ray source is shown in Figure \ref{Fig:AvClusCurrent}. As expected, a rise towards higher fluxes can be seen. While at low flux rates, both detectors have a similar behavior, we observe a larger number of reconstructed clusters for the RC concept compared to the CC detector for large flux rates. 

The average signal duration time $\Delta t$ vs. the incident flux is shown in Figure \ref{Fig:SigDurCurrent} for both detectors. While the CC detector shows nearly a constant behavior, we observe a decrease of $\Delta t$ with higher fluxes for the RC detector, which has in general lower signal duration time. This can be explained by the signal formation at the readout pads, which is achieved not only via capacitive coupling but also via a direct resistive connection between the resistive pad and the underlying readout pad.  The decay time for the resistively coupled detector decreases, as expected, with the decrease of the maximum charge for increasing incident fluxes (Figure \ref{Fig:AvChargeCurrent}). Such an effect cannot be seen for the CC detector, as the full induced charge has to spread over the whole resistive layer.

The difference in the signal duration time $\Delta t$ for both detectors explains also the dependence of the average number of clusters per recorded event on the flux rate (Figure \ref{Fig:AvClusCurrent}). While we observe a very similar behavior for the detectors for small rates, we see a larger number of reconstructed clusters for the RC detectors at higher rates. Keeping in mind that clusters are defined as localized events, it is expected to see saturation effects for both detectors at larger rates. The shorter signal duration for the RC detector leads therefore also to smaller saturation effects, as the RC detector stays more efficient. 

\section{\label{sec:concl}Conclusion}

In this paper we presented two prototype detectors based on micromegas technology with a pad readout geometry that are scalable to large dimensions. The scalability is achieved by placing the readout infrastructure on the opposite side of the active detector area. As a proof of concept of this approach, two prototype detectors have been constructed and extensively tested. A signal reconstruction efficiency of over $80\%$ has been achieved. In addition, two different implementations of a spark-resistent protection layer have been used. While the first method followed the traditional approach of a resistive protection layer that has been successfully used in previous Micromegas strip detectors, the second method is based on a novel approach: Instead of only allowing for a capacitive coupling of the resistive layer to the readout pad, we connect each resistive pad to the corresponding readout pad via a highly resistive connection. It has been shown that this approach leads to significantly more localized signals on the pad detector and smaller signal duration times, and hence improves the performance in high rate environments.  

\section*{Acknowledgements}

We would like to acknowledge the close collaboration with Rui de Oliveira from the CERN PCB workshop. In addition, we would like to thank J. Bortfeld for the help during the construction process, as well as F. Fiedler and T. Alexopoulos for the useful comments during the preparation this paper. This work was supported by the Volkswagen Foundation and the German Research Foundation (DFG).

%%%%%%%%%%%%%%%%%%%%%%%%%%
\bibliography{MicromegasPAD}{}
\bibliographystyle{atlasnote}

\end{document}